\documentclass[showpacs,amsmath,amssymb,onecolumn,pra,superscriptaddress]{revtex4-1}

\usepackage[dvips]{graphicx} 
\usepackage{amsfonts}
\usepackage{amssymb}
\usepackage{amscd}
\usepackage{amsmath}    
\usepackage{enumerate}
\usepackage{epsfig}
\usepackage{subfigure}
\usepackage{xcolor}
\usepackage{amsthm}

\newcommand{\bra}[1]{\mbox{$\left\langle #1 \right|$}}
\newcommand{\ket}[1]{\mbox{$\left| #1 \right\rangle$}}

\begin{document}
\title{Simulating single photons with realistic photon sources}

\date{\today}
\author{Xiao Yuan}
\author{Zhen Zhang}
\affiliation{Center for Quantum Information, Institute for Interdisciplinary Information Sciences, Tsinghua University, Beijing, China}
\author{Norbert L\"utkenhaus}
\affiliation{
Institute for Quantum Computing and Department of Physics and Astronomy, University of Waterloo, N2L3G1 Waterloo, Ontario, Canada}
\author{Xiongfeng Ma}
\affiliation{Center for Quantum Information, Institute for Interdisciplinary Information Sciences, Tsinghua University, Beijing, China}

\begin{abstract}
Quantum information processing provides remarkable advantages over its classical counterpart. Quantum optical systems are proved to be sufficient for realizing general quantum tasks, which however often rely on single photon sources. In practice, imperfect single photon sources, such as weak coherent state source, are used instead, which will inevitably limit the power in demonstrating quantum effects. For instance, with imperfect photon sources, the key rate of the BB84 quantum key distribution protocol will be very low, which fortunately can be resolved by utilizing the decoy state method. As a generalization, we investigate an efficient way to simulate single photons with imperfect ones to an arbitrary desired accuracy when the number of photonic inputs is small. Based on this simulator, we can thus replace the tasks that involve only a few single photon inputs with the ones that only make use of imperfect photon sources. In addition, our method also provides a quantum simulator to quantum computation based on quantum optics. In the main context, we take phase randomized coherent state as an example for analysis. A general photon source applies similarly and may provide some further advantages for certain tasks.
\end{abstract}

\maketitle

\section{Introduction}
Quantum information science develops rapidly in the last few decades. At the theoretical level, varieties of schemes are proposed to solve classical intractable problems or provide certain quantum advantages. Specifically, the Shor's factorization algorithm \cite{shor1997polynomial} indicates that quantum computing can exponentially enhance the computational power in certain tasks compared to a classical computer. In addition, quantum key distribution (QKD) protocols  \cite{BB84,Ekert:QKD:1991} enable remote users to extend secret keys with security guaranteed by basic principles of quantum mechanics.

In experiment, quantum optics is favoured for realizing quantum information processing tasks due to the weak interaction between photon and its environment. Especially in quantum communication, varieties of tasks, such as long distance quantum key distribution \cite{Ursin07}, quantum teleportation \cite{yin2012quantum} are realized with linear optics.
On the other hand, linear optics is not enough to realize universal quantum computation. Roughly speaking, it requires exponentially large resources to implement a linear quantum optical computer \cite{Cerf98}.
Thus, nonlinearity is crucial for universal quantum computation in linear optics. One possible way is to use nonlinear optics \cite{Kok:07:rmp}, which still faces the scalable difficulty for current technology. On the other hand, Knill, Laflamme and Milburn (KLM) \cite{knill2001scheme} have shown that efficient quantum computation is possible using only linear optics with single photon sources. The nonlinearity is introduced by adaptive measurements which can be realized with the techniques of quantum teleportation \cite{gottesman1999demonstrating}.


In reality, perfect single photons sources and detectors are not available. Instead, other imperfect photon sources, such as, heralded spontaneous parametric down-conversion (SPDC) sources, are used to simulate single photons. Meanwhile, single photon detectors generally have low efficiency. The imperfection of devices will lead to unexpected events, thus limiting the quantum advantage. In experiment, entangling eight photons is the best reported result \cite{yao2012observation, huang2011experimental}.

The imperfections of  single photon sources not only affect the accuracy but cause loopholes in cryptography protocols. Specifically, the multi-photon parts will lead to photon number splitting attacks \cite{BLMS:PNS:2000}, which makes the key rate of the well-known BB84 QKD protocol \cite{BB84} very low. The imperfect photon sources seem to limit the power of optical realization of quantum information processing. Surprisingly, this is not the case in reality. Even with imperfect photon sources, such as, weak coherent state as an input, secure QKD protocols are still possible by utilizing the decoy state method \cite{Hwang:Decoy:2003,Lo:Decoy:2005,Wang:Decoy:2005}. By inputting two or more coherent states, one can still estimate the information leaked in eavesdropping and thus make the whole process secure.

In this work, we generalize the idea of decoy state to general optical circuits. As an example, we show that it is possible to simulate a single photon to arbitrary accuracy efficiently by making efficient uses of phase randomized coherent states.  In addition, we generalize our result to multiple photons. We show that, replacing a few single photons with coherent states is possible in general quantum information tasks. For large numbers of photons, we link our work to the scenario of quantum computation. Our method thus provide a quantum simulator to general quantum computing processes. At last, we discuss that our method works for a general photon sources.

\section{Framework} \label{Sec:FockModel}
\begin{figure}[hbt]
\centering \resizebox{8cm}{!}{\includegraphics{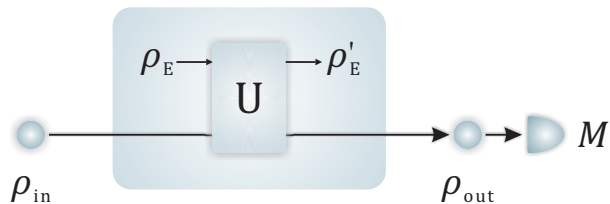}}
\caption{Optical circuits with a single photon as input.} \label{Fig:Channel}
\end{figure}

In this section, we first review the basic framework of optical circuits. With a single photon as input, whose density matrix is denoted by $\rho_{\mathrm{in}}$,  a general optical circuit can be regarded as a quantum channel described in Fig.~\ref{Fig:Channel}, which involves an unitary interaction $U$ between the signal photon and the environment $E$. After the channel, a measurement $M$ is performed on the output photon $\rho_{\mathrm{out}}$.

Such a general quantum channel can be fitted into many scenarios such as QKD, where Alice encodes her information in the input signal state $\rho_{\mathrm{in}}$ and send it through a public quantum channel to Bob.
On the output side, Bob performs photon number measurement on his received signal  $\rho_{\mathrm{out}}$, denoted by a positive-operator valued measure (POVM). For a specific POVM element $M$, the detection probability is,
\begin{equation} \label{SMCQC:POVM}
\begin{aligned}
Q = \mathrm{Tr}\left[M\rho_{\mathrm{out}}\right],
\end{aligned}
\end{equation}
where $\rho_{\mathrm{out}} = \mathrm{Tr}_E[U(\rho_{\mathrm{in}}\otimes\rho_{\mathrm{E}})U^\dag]$.
Necessary classical postprocessing should be applied to the outcome $Q$ to extract final desired quantum information.
In the following, we will focus on this specific POVM element, it is straightforward to see that our results applies similarly to the other POVM elements.

Ideally, the information is encoded on a single photon. While, a more general scenario is that Alice feeds a mixture of Fock states,
\begin{equation} \label{SMCQC:input}
\begin{aligned}
\rho_{\mathrm{in}} '= \sum_{k=0}^{\infty} P(k)\ket{k}\bra{k} \\
\end{aligned}
\end{equation}
where \ket{k} represents a Fock state that contains $k$ photons and $P(k)$ is the photon number distribution satisfying $P(k)\in[0,1]$ and $\sum_{k=0}^{\infty} P(k) = 1$. For a single photon source, $P(k)$ is a Kronecker delta function,
\begin{equation} \label{SMCQC:SinglePhoton}
\begin{aligned}
P(k) &= \delta_{k1} = \begin{cases} 1, & k=1 \\ 0, & k\neq1 \end{cases}. \\
\end{aligned}
\end{equation}
For  phase randomized coherent state source, $P(k)$ is a Poisson distribution \cite{Lo:Decoy:2005},
\begin{equation} \label{SMCQC:Poisson}
\begin{aligned}
P(k) &= \frac{\mu^k}{k!}e^{-\mu}. \\
\end{aligned}
\end{equation}

When a general photon state defined in Eq.~\eqref{SMCQC:input} is used, Bob's detection probability defined in Eq.~\eqref{SMCQC:POVM} is given by
\begin{equation} \label{}
\begin{aligned}
Q = \sum_{k=0}^{\infty}P(k)P(\mathrm{click}|k \,\mathrm{photons}),
\end{aligned}
\end{equation}
where $P(\mathrm{click}|k \,\mathrm{photons})$ denotes the detection probability with $k$ photons as input, that is,
\begin{equation}\label{}
  P(\mathrm{click}|k \,\mathrm{photons}) = \mathrm{Tr}\left[MU(\ket{k}\bra{k}\otimes\rho_{\mathrm{E}})U^\dag\right]
\end{equation}
For easier presentation, we will denote $P(\mathrm{click}|k \,\mathrm{photons})$ by $Y_k$. When $k=0$, $Y_0$ corresponds to the yield with no photon input, i.e., dark count. When $k=1$, $Y_1$ denotes the probability with single photon as input. For a single photon source, we have $Q = Y_1$; While, for a phase randomized coherent source, we have
\begin{equation} \label{SMCQC:Qone}
\begin{aligned}
Q = \sum_{k=0}^{\infty}\frac{\mu^k}{k!}e^{-\mu}Y_k.
\end{aligned}
\end{equation}

Ideally, single photon source is required for several quantum information processing tasks. The accurate probability distribution $Y_1$ can only be obtained with single photon source. However, if we are only intended in learning the value of $Y_k$, we show in this work that such value can be accurately and efficiently estimated with several phase randomized coherent states as input.

\section{Simulating single photon} \label{Sec:ParaEstSingle}
In this section, we will focus on simulating the probability distribution $Y_1$ with a single photon of a general quantum circuit.
For easier presentation, let us first define $A_\mu = Q e^\mu$, that is,
\begin{equation} \label{SMCQC:Aone}
\begin{aligned}
A_\mu = \sum_{k=0}^{\infty}\frac{\mu^k}{k!}Y_k.
\end{aligned}
\end{equation}
To estimate $Y_1$, we will make use of the the idea of decoy-state method originally applied in QKD \cite{Ma05}. That is, Alice chooses a few probe intensities of phase randomized coherent states to get several detections $A_\mu$. By regarding the probability $Y_i$ with $i$ photons, for $i = 1, 2, \dots$, as unknown variables, we thus get several linear equations of $Y_i$ in the form of Eq.~\eqref{SMCQC:Aone} with different $\mu$ and $A_{\mu}$. As there are infinite numbers of unknown variables, we need infinite numbers of equations to deterministically decide $Y_1$. While, we can still approximately estimate $Y_1$ with finite linear equations. With more number of coherent states used, the estimation becomes more accurate. Similar analysis has been done for QKD with a few decoy states \cite{harrington2005enhancing}, however, the obtained estimation is not optimal.

For heuristic presentation, we will show examples of estimating $Y_1$ with one and two probe intensities. Another example with three probe intensities can be found in Appendix.~\ref{Sec:3probe}. Furthermore, we analytically derive an estimator for $Y_1$ with general $L$ probe intensities plus one vacuum intensity. With an explicit example, we numerically prove that the estimation error decays exponentially with the number of probes $L$. These results are derived in the asymptotic case, where for each probe intensity there are infinite number of samples. To take finite size effect into account, we consider finite number of samples $M/(L+1)$ for each probe intensity. By considering the total error of our method including estimation and statistical error, we find that it scales inversely proportional to a power function of the total number usages of coherent pulses $M$, $\exp(-O(\log(M)))$. Therefore, our method with coherent probes is efficient.

In the following, we will first discuss the case with one, two and three probe intensities, and then generalize the result to $L$ probe intensities. The discussion of one, two and three probe intensities   can be found in the decoy state method in QKD \cite{Ma05} and generalization to $L$ probe intensities is our new result.
\subsection{Review: one probe intensity}
Firstly, we consider that only one phase randomized coherent state $\rho_\mu$ is used. In this case, an estimation of $Y_1$ is given by the redefined probability $A_\mu$ in Eq.~\eqref{SMCQC:Aone} divided by its intensity $\mu$, that is, $Y_1^{\mathrm{est}} = A_\mu/\mu$. To see the estimation accuracy, we use the relation between $A_\mu$ and $Y_1$ via Eq.~\eqref{SMCQC:Aone},
\begin{equation} \label{SMCQC:Aone1}
\begin{aligned}
Y_1 &= \frac{1}{\mu}\left(A_\mu - Y_0 - \frac{\mu^2}{2}Y_2 - \dots\right) \\
&= Y_1^{\mathrm{est}} - \frac{1}{\mu}\left(Y_0 + \frac{\mu^2}{2}Y_2 + \dots\right)
\end{aligned}
\end{equation}
As $Y_n$ corresponds to the probability with $n$ photons as input, we have $Y_n\in[0,1]$ and hence the bounds of $Y_1$,
\begin{equation} \label{SMCQC:Aone1bounds}
\begin{aligned}
Y_1^{\mathrm{est}} - \frac{e^\mu-\mu}{\mu} \le Y_1 \le Y_1^{\mathrm{est}}
\end{aligned}
\end{equation}
The estimation accuracy is defined by the interval between the upper and lower bounds
\begin{equation} \label{SMCQC:oneDelta0}
\begin{aligned}
\Delta_0 = \frac{e^\mu-\mu}{\mu}
\end{aligned}
\end{equation}
which is minimized at $\mu = 1$ with the value of $e-1>1$. Thus, at least one of the bounds in Eq.~\eqref{SMCQC:Aone1bounds} is trivial since $Y_1\in[0,1]$.

It is easy to see that a single use of coherent state gives very loose estimation of $Y_1$. This can be intuitively understood by the dark count contribution $Y_0$ appeared in Eq.~\eqref{SMCQC:Aone1}. To overcome this, we can input an additional coherent state and show in the following that $Y_1$ can be estimated to a much better accuracy.

\subsection{Review: vacuum + one probe intensities} \label{Sub:OneV1}
Now, suppose Alice can add another probe coherent state $\rho_\nu$. In this scenario, there are two linear equations,
\begin{equation} \label{SMCQC:one2A}
\begin{aligned}
A_\mu &= Y_0 + \mu Y_1 + \frac{\mu^2}{2}Y_2 + \dots \\
A_\nu &= Y_0 + \nu Y_1 + \frac{\nu^2}{2}Y_2 + \dots
\end{aligned}
\end{equation}
Subtract one from the other, we have
\begin{equation} \label{SMCQC:oneFinEq1}
\begin{aligned}
\frac{A_\mu-A_\nu}{\mu-\nu} &= Y_1 +\frac{\mu+\nu}{2}Y_2 +\dots. \\
\end{aligned}
\end{equation}
Therefore, we can estimate $Y_1$ by $Y_1^{\mathrm{est}} = \frac{A_\mu-A_\nu}{\mu-\nu}$, and have a relation
\begin{equation} \label{SMCQC:oneFinEq2}
Y_1 = Y_1^{\mathrm{est}} -\frac{\mu+\nu}{2}Y_2-\dots.
\end{equation}

Assume $\mu>\nu$, the bounds of $Y_1$ are given by
\begin{equation} \label{SMCQC:one2Y1bds}
\begin{aligned}
Y_1^{\mathrm{est}}  - \frac{e^\mu - \mu - e^\nu + \nu}{\mu-\nu} \le Y_1 \le Y_1^{\mathrm{est}}
\end{aligned}
\end{equation}
The size of the interval is
\begin{equation} \label{SMCQC:oneDelta1}
\begin{aligned}
\Delta_1 = \frac{e^\mu - e^\nu}{\mu-\nu} - 1
\end{aligned}
\end{equation}
The minimum of $\Delta_1$ is reached for $v = 0$ and it increases with $\mu$. For a small $\mu$, we can approximate the interval by
\begin{equation} \label{SMCQC:oneDelta1App}
\begin{aligned}
\Delta_1 = \frac{\mu}{2} + O\left(\frac{\mu^2}{2!}\right)
\end{aligned}
\end{equation}
The intuition behind the choices of the intensities comes from the motivation to estimate  the background contribution $Y_0$. After then, the estimation error of $Y_1$ suffers only from contributions of more than two photon numbers, that is, $O({\mu^2}/{2!})$. Therefore, in the following, we will always consider the vacuum probe intensity.


\subsection{Vacuum + $L$ probe intensities}
We leave the result with vacuum + 2 probe intensities in Appendix \ref{Sec:3probe}, and consider a general case that Alice inputs $L+1$ phase randomized coherent state $\rho_{\mu_0}$ (Vacuum), $\rho_{\mu_1}, \dots, \rho_{\mu_L}$. Suppose $\mu_0=0$ and $\mu_1<\mu_2<\dots<\mu_L$, an estimation of $Y_1$ is given in Appendix~\ref{App:LQuizzes}, Eq.~\eqref{Eq:appestimate}, by
\begin{equation}\label{Eq:vacuumplusLquizzes}
  Y_1^{\mathrm{est}} = \mu_1\mu_2\cdots\mu_L\sum_{j=1}^{L}\frac{\mu_j^{-2} (A_{\mu_j}-A_0)}{\prod_{1\le n\le L;n\ne j}(\mu_n-\mu_j)},
\end{equation}
where $A_{\mu_j}$ is the gain for coherent state input with intensity $\mu_j$.
The bounds of the $Y_1$ estimation is
\begin{equation} \label{SMCQC:Y1Bounds}
\begin{aligned}
&Y_Y^{\mathrm{est}} - \Delta_L\leq Y_1\leq Y_1^{\mathrm{est}}, \textrm{for $L$ to be even}\\
&Y_1^{\mathrm{est}}\leq Y_1\leq Y_1^{\mathrm{est}} + \Delta_L, \textrm{for $L$ to be odd}
\end{aligned}
\end{equation}
where the interval between the upper and lower bounds is given according to Eq.~\eqref{Eq:appDelta} by
\begin{equation} \label{SMCQC:Delta1Lfin}
\begin{aligned}
\Delta_L
 &= (-1)^{L+1}\left(\mu_1\mu_2\dots\mu_L\sum_{j=1}^{L}\frac{\mu_j^{-2} (e^{\mu_j}-1)}{\prod_{1\le n\le L;n\ne j}(\mu_n-\mu_j)}-1\right)\\
&= \frac{\mu_1\dots\mu_L}{(L+1)!}+ O\left[\frac{\mu_1\dots\mu_L\sum\mu_l}{(L+2)!}\right].
\end{aligned}
\end{equation}
When the intensities $\mu_j$ are small, we can see that the estimation interval exponentially decreases with $L$. Thus, a single photon can be efficiently simulated with coherent source as input

According to Eq.~\eqref{Eq:y1estimation} and Eq.~\eqref{SMCQC:deltaLestimation}, the estimation $Y_1^{\mathrm{est}}$ and the interval $\Delta_L$ can be represented as a linear combination of $A_{\mu_j}$ as,
\begin{equation}\label{Eq:resultsinglemode}
\begin{aligned}
Y_1^{\textrm{est}} &=\sum_{j=1}^{\lceil L/2\rceil} \lambda_{2j-1}A_{\mu_{2j-1}} - \sum_{j=1}^{\lfloor L/2\rfloor}\lambda_{2j}A_{\mu_{2j}}+\lambda_0A_0,\\
\Delta_L  &=(-1)^{L+1}\left(\sum_{j=1}^{\lceil L/2\rceil} \lambda_{2j-1}e^{\mu_{2j-1}} - \sum_{j=1}^{\lfloor L/2\rfloor}\lambda_{2j}e^{\mu_{2j}}+\lambda_0-1\right),
\end{aligned}
\end{equation}
where the  coefficients $\lambda_j$ are positive and given by
\begin{equation}\label{Eq:resultdeflambda}
\begin{aligned}
\lambda_{0} &= \sum_{j=1}^L(-1)^j\lambda_j, \\
\lambda_{j} &= \frac{(-1)^{j+1}}{\mu_j}\prod_{1\le n\le L;n\ne j} \frac{\mu_n}{(\mu_n-\mu_j)}, \quad \text{for} \quad 1\le j\le L.
\end{aligned}
\end{equation}

We refer to Appendix~\ref{App:LQuizzes} for the derivation of the results and focus on the performance.
\subsection{Total error of estimation}\label{Sec:totalerror}
In the estimation of $Y_1$ given in Eq.~\eqref{Eq:vacuumplusLquizzes}, we assume $A_{\mu_j}$ to be accurate. In practice, we have to input several copies of the same coherent state with intensity $\mu_j$, and $A_{\mu_j}$ can be estimated from the measurement. In this case, beside the estimation error $\Delta_L$, we have to consider statistical error of estimating each $A_{\mu_j}$. In the last part, we have proved that the estimation with a few probes intensities can efficiently simulate the result with with single photon state in the asymptotical scenario. In the following, we will show that such a method is also efficient when focusing on finite data size.

To show the method to be as efficient as the with one with single photon, we consider independent and identically distributed (i.i.d.) sampling for simplicity. In QKD, such finite size effects without assumptions has been analyzed for the vacuum plus weak decoy state formalism \cite{curty2014finite, Xu14,Zhen}.  In Ref.~\cite{Zhen}, it has been shown that the difference is only a factor when the sample size is large. Thus, we leave the analysis without additional assumptions in future work.

Under the i.i.d. assumption, the statistical error $\Delta_s(A)$ of $A_{\mu}$ can be approximated by
\begin{equation}\label{}
  \Delta_s(A) \lesssim \frac{1}{\sqrt{m}} = \sqrt{\frac{L+1}{M}},
\end{equation}
where $m$ is the number of samples for each $\mu_j$, $M = m(L+1)$ is the total number of samples. Here, we consider the same statistical error estimation for all $A_{\mu}$ for simplicity. Tighter bound that involves $A_{\mu}$ can be further applied when the value of $A_{\mu}$ is known.

The sampling induced error of $Y_1^{\textrm{est}}$ is given by
\begin{equation}\label{}
\begin{aligned}
  \Delta_s(Y_1^{\textrm{est}}) = \Delta_s(A)\sqrt{\sum_{j = 0}^L \lambda_j^2}.
  \end{aligned}
\end{equation}
And the total error in experiment is thus,
\begin{equation}\label{Eq:totalerror}
\begin{aligned}
\Delta_t &\lesssim \Delta_s + \Delta_L. \\
\end{aligned}
\end{equation}

In the following, we will give an example to show the scale of total error $\Delta_t$ compared to a fixed total sample size $M$. With perfect single photon source, the total error scales as $O(1/\sqrt{M})$. With phase randomized coherent state, we show that by inputting appropriate quiz states, the total error also scales as a power function of $M$.

\subsection{Example}\label{Sec:totalerrorexample}
\begin{figure}[hbt]
\centering \resizebox{9cm}{!}{\includegraphics{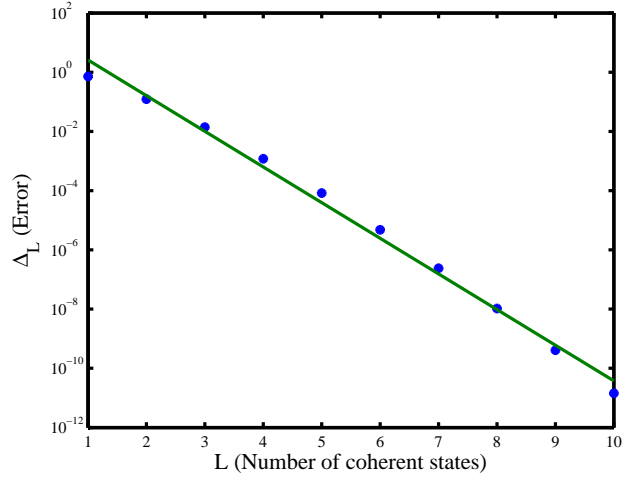}}
\caption{Error $\Delta_L$ for specific usage of coherent intensities, $\mu_j = j/L$ for $j = 0, 1, \dots, L$. $L$ runs only from 1 to 10 for computational accuracy limit. Blue dots are the $\Delta_L$ value, and the green line is the exponential fit.}\label{Fig:DeltaL}
\end{figure}

As different probe intensities will lead to different estimation error $\Delta_L$. We only take an example with probe intensities, $\mu_j = j/L$ for $j = 0, 1, \dots, L$. There may exist better choices of the probe intensities that causes a smaller total error. In our example, the coefficients $\lambda_j$, defined in Eq.~\eqref{Eq:resultdeflambda}, can be calculated by
\begin{equation}\label{}
\begin{aligned}
  \lambda_{j} &= \frac{(-1)^{j+1}}{j/L}\prod_{1\le n\le L;n\ne j} \frac{n/L}{(n/L-j/L)} \\
  &= \frac{L(-1)^{j+1}}{j}\prod_{1\le n\le L;n\ne j} \frac{n}{(n-j)}\\
  &=\frac{L}{j} {L\choose j},
  \end{aligned}
\end{equation}
and $\lambda_0$ by
\begin{equation}\label{}
\begin{aligned}
  \lambda_{0} &= \sum_{j=1}^L(-1)^j\frac{L}{j} {L\choose j},
  \end{aligned}
\end{equation}
In this case, the estimation error $\Delta_L$ can be numerically calculated as shown in Fig.~\ref{Fig:DeltaL}. A linear fitting between $\log{\Delta_L}$ and $L$ thus gives the relation $\log{\Delta_L} = -2.772 * L + 3.718$.
It is straightforward to see that with increasing $L$, the interval $\Delta_L$ exponentially approaches 0. 

The sampling error is
\begin{equation}\label{}
\begin{aligned}
  \Delta_s(Y_1^{\textrm{est}}) &= \Delta_s(A)f(L),
  \end{aligned}
\end{equation}
where $f(L)$ is a constant factor
\begin{equation}\label{}
f(L) = \sqrt{\left(\sum_{j=1}^L(-1)^j\frac{L}{j} {L\choose j}\right)^2 + \sum_{j=1}^L \left(\frac{L}{j} {L\choose j}\right)^2}.
\end{equation}

\begin{figure}[hbt]
\centering \resizebox{9cm}{!}{\includegraphics{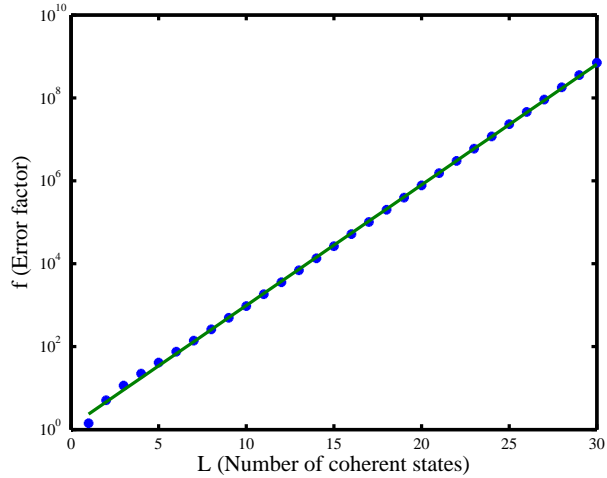}}
\caption{Error factor $f$ for different $L$. Blue dots are the $\Delta_L$ value, and the green line is the exponential fit.}\label{Fig:error}
\end{figure}

As shown in Fig.~\ref{Fig:error}, the error factor $f$ is roughly exponential to the number of coherent states $L$. A linear fitting thus gives the relation $\log{f} = 0.67L + 0.189$.

\begin{figure}[hbt]
\centering \resizebox{9cm}{!}{\includegraphics{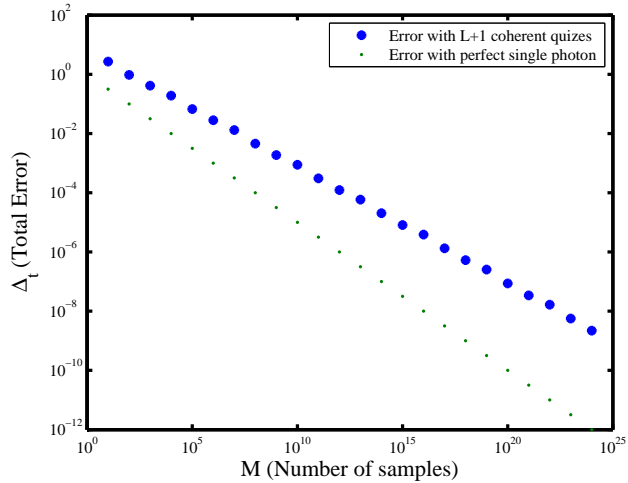}}
\caption{Optimized total error for different number of samples. The Blue dots are the total error for $L(M)$ coherent probe intensities. The green dots are the total error for single photon source, that is, $1/\sqrt{M}$.}\label{Fig:DeltaTM}
\end{figure}

Thus, the total error can be approximated modeled by
\begin{equation}\label{}
  \Delta_t = \sqrt{\frac{L+1}{M}}e^{0.67L + 0.189} + e^{-2.772L + 3.718}.
\end{equation}
For given total number of samples $M$, we can optimize over $L$ to minimize the total error $\Delta_t$. We solve this problem numerically. As shown in Fig.~\ref{Fig:DeltaTM}, the total error $\Delta_t$ is still inversely proportional to a power function of the number of samples $M$. That is, we have $\Delta_t \approx 6.6128/M^{0.3931}$ which fits the data we present in Fig.~\ref{Fig:DeltaTM}.

\begin{figure}[hbt]
\centering \resizebox{9cm}{!}{\includegraphics{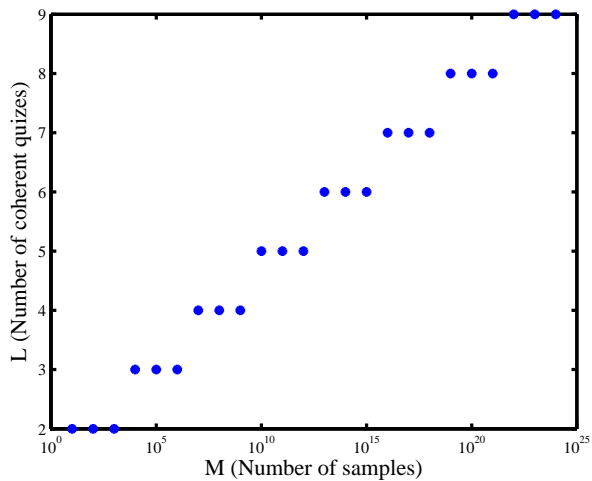}}
\caption{Number of non-zero probe intensities $L$ for different number of samples $M$. }\label{Fig:Quizes}
\end{figure}

In addition, the optimized number of probe intensities is shown in Fig.~\ref{Fig:Quizes}. Roughly speaking, $L$ is linearly proportional to $\log{M}$, which explains why $\Delta_t$ is still a power function of $M$.

\section{Parameter estimation: multiple input modes} \label{Sec:ParaEstMulti}
\begin{figure}[hbt]
\centering \resizebox{8cm}{!}{\includegraphics{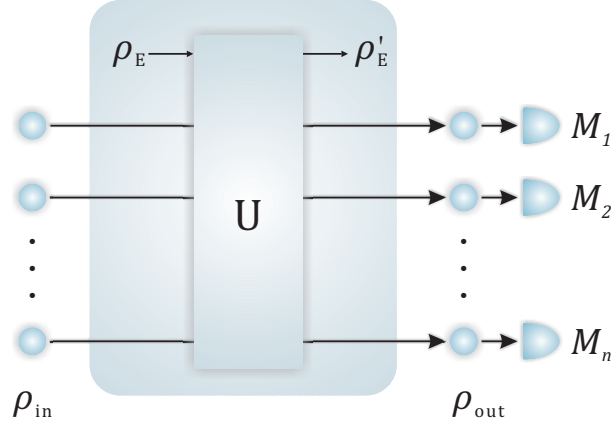}}
\caption{A schematic diagram for the a general quantum channel with $n$ optical input modes.} \label{Fig:MultiChannel}
\end{figure}

Now, we consider a general optical circuits with $n$ distinguishable photons as inputs. The quantum circuits can be well described by a quantum channel with $n$ optical modes as shown in Fig.~\ref{Fig:MultiChannel}. The input state $\rho_{\mathrm{in}}$ consists of $n$ single photons which corresponds to each of the input mode. After a unitary interaction between the input particles and the environment described by $\rho_E$, measurements are performed on each of the output mode.
For each of the measurement $M_i$, for $i = 1, 2, \dots, n$, the detection probability is given by,
\begin{equation} \label{}
\begin{aligned}
Q_i = \mathrm{Tr}_{\neq i, E}\left[M_i\rho_{\mathrm{out}}\right] = \mathrm{Tr}_{\neq i, E}\left[M_iU(\rho_{\mathrm{in}}\otimes\rho_{\mathrm{E}})U^\dag \right],  \\
\end{aligned}
\end{equation}
where the trace is over the environment $E$ and all the input modes except the $i$th one.

In the previous section, we show that each single mode photon can be simulated efficiently with multiple usages of coherent pulses.
Here, we generalize the result to $n$ input modes case. We define a coincidence detection by
\begin{equation} \label{}
\begin{aligned}
Q = \mathrm{Tr}\left[M\rho_{\mathrm{out}}\right]= \mathrm{Tr}\left[MU(\rho_{\mathrm{in}}\otimes\rho_{\mathrm{E}})U^\dag \right],
\end{aligned}
\end{equation}
where the measurement is $M = M_1\otimes M_2\otimes \cdots \otimes M_n$.
When the input state is a mixture of photon number states,
\begin{equation} \label{SMCQC:input2}
\begin{aligned}
\rho_{\mathrm{in}} = \sum_{k_1,k_2,\dots,k_n=0}^{\infty}  P(k_1,k_2,\dots, k_n)\ket{k_1k_2\dots k_n}\bra{k_1k_2\dots k_n}, \\
\end{aligned}
\end{equation}
the coincidence detection can be expressed by
\begin{equation} \label{}
\begin{aligned}
Q = \sum_{k_1,k_2,\dots,k_n=0}^{\infty}  P(k_1,k_2,\dots, k_n)Y_{k_1k_2\dots k_n},
\end{aligned}
\end{equation}
where $Y_{k_1k_2\dots k_n}$ is the coincidence detection probability for the case that the $i$th mode has $k_i$ number of photons,
\begin{equation} \label{}
\begin{aligned}
Y_{k_1k_2\dots k_n} = \mathrm{Tr}\left[MU(\ket{k_1k_2\dots k_n}\bra{k_1k_2\dots k_n}\otimes\rho_{\mathrm{E}})U^\dag \right].
\end{aligned}
\end{equation}

In the following, we show that with coherent state as input, we can also estimate the coincidence detection for single photon input, $Y_{11\dots1}$, to an arbitrary accuracy. For simplicity, we consider the same probe intensities for different input modes. The derivation of different probe intensities for different input modes follows similarly.

\subsection{Two modes with vacuum + one probe intensities}
Firstly, we consider only two input optical modes and two probe intensities. From Section \ref{Sub:OneV1}, we find that one of the two probe intensities should be a vacuum state and the other should be a weak state $\rho_\mu$ in the optimal case. Similar to Eq.~\eqref{SMCQC:Qone}, we have
\begin{equation} \label{SMCQC:Qtwo2}
\begin{aligned}
Q = \sum_{k_1=0}^{\infty}\frac{\mu_1^{k_1}}{{k_1}!}e^{-\mu_1} \sum_{k_2=0}^{\infty}\frac{\mu_2^{k_2}}{{k_2}!}e^{-\mu_2}Y_{k_1k_2},
\end{aligned}
\end{equation}
where $\mu_1$ ($\mu_2$) and $k_1$ ($k_2$) are the coherent state intensity and the photon number for the 1 (2) mode, respectively
Similar to Eq.~\eqref{SMCQC:Aone}, we define $A_{\mu_1\mu_2} = Q e^{\mu_1} e^{\mu_2}$,
\begin{equation} \label{SMCQC:Atwo2}
\begin{aligned}
A_{\mu_1\mu_2} = \sum_{k_1,k_2=0}^{\infty}\frac{\mu_1^{k_1}\mu_2^{k_2}}{k_1!k_2!}Y_{k_1k_2}.
\end{aligned}
\end{equation}

When each mode is input with coherent state with $0$ and $\mu$ intensity,
we have four equalities based on Alice's four possible input cases,
\begin{equation} \label{SMCQC:two2A}
\begin{aligned}
A_{00} &= Y_{00} \\
A_{\mu0} &= \sum_{k_1=0}^{\infty}\frac{\mu^k_1}{k_1!}Y_{k_10} = Y_{00} + \mu Y_{10} + \frac{\mu^2}{2}Y_{20} + \dots \\
A_{0\mu} &= \sum_{k_2=0}^{\infty}\frac{\mu^k_2}{k_2!}Y_{0k_2} = Y_{00} + \mu Y_{01} + \frac{\mu^2}{2}Y_{02} + \dots \\
A_{\mu\mu} &= \sum_{k_1,k_2=0}^{\infty}\frac{\mu^k_1\mu^{k'}}{k_1!k_2!}Y_{k_1k_2} = Y_{00} + \mu Y_{10} + \mu Y_{01} + \mu^2 Y_{11} + \dots \\
\end{aligned}
\end{equation}
With the attempt to estimate $Y_{11}$, we can linearly combine $Y_{00}$, $Y_{10}$, $Y_{01}$ and $Y_{11}$ by
\begin{equation} \label{SMCQC:two2EqA}
\begin{aligned}
Y_{11}^{\mathrm{est}} &= A_{\mu\mu} - A_{0\mu} - A_{\mu0} + A_{00} \\
&= \sum_{k,k'=0}^{\infty}\frac{\mu^k\mu^{k'}}{k!k'!}Y_{kk'} - \sum_{k=0}^{\infty}\frac{\mu^k}{k!}Y_{k0} - \sum_{k=0}^{\infty}\frac{\mu^k}{k!}Y_{0k} + Y_{00} \\
&= \sum_{k,k'=1}^{\infty}\frac{\mu^k\mu^{k'}}{k!k'!}Y_{kk'}
\end{aligned}
\end{equation}
As the $Y$s are always in $[0,1]$, we can bound $Y_{11}$ by
\begin{equation} \label{SMCQC:two2Y11}
\begin{aligned}
Y_{11}^{\mathrm{est}}- \Delta_{1, 2} \le Y_{11} \le Y_{11}^{\mathrm{est}}
\end{aligned}
\end{equation}
with a size of the interval $\Delta_{1, 2}$
\begin{equation} \label{SMCQC:two2Delta11}
\begin{aligned}
\Delta_{1, 2} = \frac{1}{\mu^2}\sum_{n,m = 1}^{\infty}\frac{\mu^n\mu^m}{n!m!} - 1= \frac{(e^\mu - 1)^2}{\mu^2} - 1
\end{aligned}
\end{equation}
Here, the first subscript denotes the number of non-zero probe intensities and the second subscript denotes the number of input modes. For a small $\mu$, we have
\begin{equation} \label{SMCQC:two2Delta11App}
\begin{aligned}
\Delta_{1, 2} = \mu + O(\mu^2)
\end{aligned}
\end{equation}

\subsection{$n$ modes with vacuum + one probe intensities}
Then, we generalize the result to the case of $n$ input modes each with two possible (vacuum $0$ and weak $\mu$) probe intensities.
Denote the nonzero coherent state intensity for the $i$th mode by $\mu_i$, the measurement result is given by, similar to Eq.~\eqref{SMCQC:Qtwo2},
\begin{equation} \label{SMCQC:Qn}
\begin{aligned}
Q = \sum_{k_1,k_2,\dots,k_n=0}^{\infty} \frac{\mu_1^{k_1}\mu_2^{k_2}\dots\mu_n^{k_n}}{k_1!k_2!\dots k_n!}e^{-(\mu_1+\mu_2+\cdots+\mu_n)}Y_{k_1k_2\dots k_n}
\end{aligned}
\end{equation}
Similar to Eq.~\eqref{SMCQC:Atwo2}, define $A_{\mu\mu\dots\mu} = Q e^{(\mu_1+\mu_2+\cdots+\mu_n)}$, and we have
\begin{equation} \label{SMCQC:An}
\begin{aligned}
A_{\mu_1\mu_2\dots\mu_n} = \sum_{k_1,k_2,\dots,k_n=0}^{\infty} \frac{\mu_1^{k_1}\mu_2^{k_2}\dots\mu_n^{k_n}}{k_1!k_2!\dots k_n!}Y_{k_1k_2\dots k_n}.
\end{aligned}
\end{equation}

For easier presentation, we first introduce an operation on the coincidence probability $Y$. For $Y_{k_1k_2\dots k_n}$, we define it to be
\begin{equation}\label{}
  Y_{k_1k_2\dots k_n} = \bigotimes_{i=1}^{n} Y_{k_i},
\end{equation}
where the operation $\bigotimes_{i=1}^{n}$ denotes subscript combination. The notation $Y_{k_i}$ does not make sense unless the  the operation $\bigotimes_{i=1}^{n}$ is applied. With this notation, we can rewrite $A_{\mu_1\mu_2\dots\mu_n}$ by
\begin{equation} \label{SMCQC:defsubk}
\begin{aligned}
A_{\mu_1\mu_2\dots\mu_n} =
\bigotimes_{i=1}^{n}A_{\mu_i}
\end{aligned}
\end{equation}
where
\begin{equation} \label{SMCQC:defsubk2}
\begin{aligned}
A_{\mu_i} &= \sum_{k_i=0}^\infty\frac{\mu_i^{k_i}}{k_i!}Y_{k_i}
\end{aligned}
\end{equation}
Notice that, $\bigotimes_{i=1}^{n}$ can still be regarded as a product operation where the multiplication of the $Y$s is replaced with subscript combination. Thus,
\begin{equation} \label{SMCQC:defsubk3}
\begin{aligned}
\bigotimes_{i=1}^{n}A_{\mu_i}&=  \sum_{k_1,k_2,\dots,k_n=0}^{\infty} \frac{\mu_1^{k_1}\mu_2^{k_2}\dots\mu_n^{k_n}}{k_1!k_2!\dots k_n!}\bigotimes_{i=1}^{n} Y_{k_i} \\
&= A_{\mu_1\mu_2\dots\mu_n} \\
\end{aligned}
\end{equation}

With the same spirit, we can derive
\begin{equation} \label{SMCQC:n2defk}
\begin{aligned}
\bigotimes_{i=1}^{n}\left(A_{\mu_i} - A_0\right) &= \bigotimes_{i=1}^{n}\left(\sum_{k_i=0}^\infty\frac{\mu_i^{k_i}}{k_i!}Y_{k_i} - Y_{k_i=0}\right)\\
&=\sum_{k_1,k_2,\dots,k_n=1}^{\infty} \frac{\mu^{k_1}\mu^{k_2}\dots\mu^{k_n}}{k_1!k_2!\dots k_n!}Y_{k_1k_2\dots k_n},
\end{aligned}
\end{equation}
which is a generalization to Eq.~\eqref{SMCQC:two2EqA}. Here, we let all $\mu_i$ equal to the same intensity $\mu$. Similar to Eq.~\eqref{SMCQC:two2EqA}, it is not hard to see that an estimation of $Y_{11\dots1}$ is given by
\begin{equation}\label{}
  Y_{11\dots1}^{\mathrm{est}} = \bigotimes_{i=1}^{n}\left(A_{\mu_i} - A_0\right).
\end{equation}
Now, the size of the interval of estimating $Y_{11\dots1}$ is given by
\begin{equation} \label{SMCQC:Deltan2}
\begin{aligned}
\Delta_{1, n} &= \frac{1}{\mu^n} \sum_{k_1,\dots k_n=1}^{\infty} \frac{\mu^{k_1}\dots\mu^{k_n}}{k_1!\dots k_n!} -1 \\
&= \frac{(e^\mu - 1)^n}{\mu^n}-1 \\
&= \frac{n}{2}\mu + O(\mu^2) \\
\end{aligned}
\end{equation}
which is consistent with Eqs.~\eqref{SMCQC:two2Delta11} and \eqref{SMCQC:two2Delta11App}.

\subsection{$n$ modes with vacuum + $L$ probe intensities}
Now, we show an estimation of $Y_{11\dots1}$ in the case that each mode is input with vacuum + $L$ probe intensities.
For each mode, the estimation can be given according to Eq.~\eqref{Eq:resultsinglemode}. Follow a similar way in the last two section, we can similarly define the $n$ mode estimation $Y_{11\dots1}^{\mathrm{est}}$ of $Y_{11\dots1}$ according to
\begin{equation}\label{eq:Alestn}
\begin{aligned}
Y_{11\dots1}^{\mathrm{est}} &= \bigotimes\left(\sum_{j=1}^{\lceil L/2\rceil} \lambda_{2j-1}A_{\mu_{2j-1}} - \sum_{j=1}^{\lfloor L/2\rfloor}\lambda_{2j}A_{\mu_{2j}}+\lambda_0A_0\right), \\
\end{aligned}
\end{equation}
where $\lambda_j$ is defined in Eq.~\eqref{Eq:resultdeflambda}. Here the product $\bigotimes$ denotes a multiplication of $A$ that is define in Eq.~\eqref{SMCQC:defsubk}. Then the estimation interval is
\begin{equation}\label{Eq:estierrorLN}
\begin{aligned}
  \Delta_{L, n} &= \left|\left(\mu_1\mu_2\dots\mu_L\sum_{j=1}^{L}\frac{\mu_j^{-2} (e^{\mu_j}-1)}{\prod_{1\le n\le L;n\ne j}(\mu_n-\mu_j)}\right)^n-1\right|\\
  &= \left|\left((-1)^{L+1}\frac{\mu_1\dots\mu_L}{(L+1)!}+ O\left[\frac{\mu_1\dots\mu_L\sum\mu_l}{(L+2)!}\right] + 1\right)^n-1\right|\\
  &= \frac{n\mu_1\dots\mu_L}{(L+1)!}+ O\left[\frac{n\mu_1\dots\mu_L\sum\mu_l}{(L+2)!}\right].
\end{aligned}
\end{equation}
Compared to the estimation error of a single photon given in Eq.~\eqref{SMCQC:Delta1Lfin}, we can see that an extra factor $n$ is added when simulating $n$ photons. As the estimation error for a single photon decays exponentially to $L$, the estimation for $n$ photons is still efficient.

In practice, to get $Y_{11\dots1}^{\mathrm{est}}$, one has to get $A_{\mu_1\mu_2\dots\mu_n}$. Suppose, for each mode, there are $L$ probe intensities used, then there are $L^n$ different number of values $A_{\mu_1\mu_2\dots\mu_n}$ to be measured. For small number of $n$, we can see that the estimation is efficient and accurate. However, the total number of probes scales exponentially with $n$. Therefore, simulating large number single photons with phase randomized coherent state is not efficient.

\section{Total error of estimation}
In Sec.~\ref{Sec:totalerror} and \ref{Sec:totalerrorexample}, we show the total error of the estimation when considering finite sample size. In general, the total error with $n$ input modes consists of the estimation error $\Delta_{L, n}$ and the statistical error $\Delta_{s,n}$,
\begin{equation}\label{Eq:totalerror2}
\begin{aligned}
  \Delta_{t,n} &\approx \Delta_{s,n} + \Delta_{L,n}. \\
  \end{aligned}
\end{equation}
The estimation error $\Delta_{L, n}$ is given in Eq.~\eqref{Eq:estierrorLN}. When $\Delta_L$ is small enough and $n$ is not large, we can approximate $\Delta_{L, n}$ by
\begin{equation}\label{}
\Delta_{L, n} = n\Delta_{L,1}.
\end{equation}
The statistical error $\Delta_{s,n}$ consists of statistical fluctuation when estimating  $A_{\mu_1\mu_2\dots\mu_n}$ for different probe intensities $\{\mu_1\mu_2\dots\mu_n\}$.  Similar to the case with one input mode, we consider the same statistical error for all $A_{\mu_1\mu_2\dots\mu_n}$ by
\begin{equation}\label{}
  \Delta_{s,n}(A_{\mu_1\mu_2\dots\mu_n}) = \lesssim \frac{1}{\sqrt{m}} = \sqrt{\frac{(L+1)^n}{M}},
\end{equation}
where $M$ denotes the total number of samples. Note that the estimation $Y_{11\dots1}^{\mathrm{est}}$ given in Eq.~\eqref{eq:Alestn} can be reformulated by
\begin{equation}\label{}
  Y_{11\dots1}^{\mathrm{est}} = \sum_{j_1, j_2,\dots,j_n = 0}^{L}\lambda_{j_1, j_2,\dots,j_n}A_{\mu_{j_1}\mu_{j_2}\dots\mu_{j_n}},
\end{equation}
where $\lambda_{j_1, j_2,\dots,j_n} = \lambda_{j_1}\lambda_{j_2}\cdots\lambda_{j_n}$
In this case, the sample error of $Y_{11\dots1}^{\mathrm{est}}$ can be given by
\begin{equation}\label{}
\begin{aligned}
  \Delta_{s,n}(Y_{11\dots1}^{\mathrm{est}}) &= \Delta_{s,n}(A_{\mu_{j_1}\mu_{j_2}\dots\mu_{j_n}})f(L,n)
\end{aligned}
\end{equation}
where
\begin{equation}\label{}
\begin{aligned}
f(L,n) &= \sqrt{\sum_{j_1, j_2,\dots,j_n = 0}^{L}\lambda_{j_1, j_2,\dots,j_n}^2}\\
&= \sqrt{\sum_{j_1, j_2,\dots,j_n = 0}^{L}\lambda_{j_1}^2\lambda_{j_2}^2\cdots\lambda_{j_n}^2}\\
&= \sqrt{\sum_{j_1 = 0}^{L}\lambda_{j_1}^2\sum_{ j_2 = 0}^{L}\lambda_{j_2}^2\cdots\sum_{j_n= 0}^{L}\lambda_{j_n}^2}\\
&= f(L,1)^n.
\end{aligned}
\end{equation}

\begin{figure}[hbt]
\centering \resizebox{9cm}{!}{\includegraphics{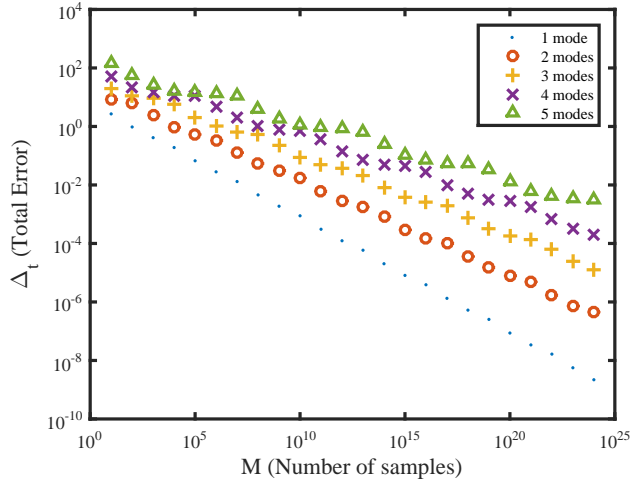}}
\caption{Optimized total error for different number of samples and different input modes. }\label{Fig:TotalerrorParty}
\end{figure}

\begin{figure*}[hbt]
\centering
\resizebox{20cm}{!}{\includegraphics{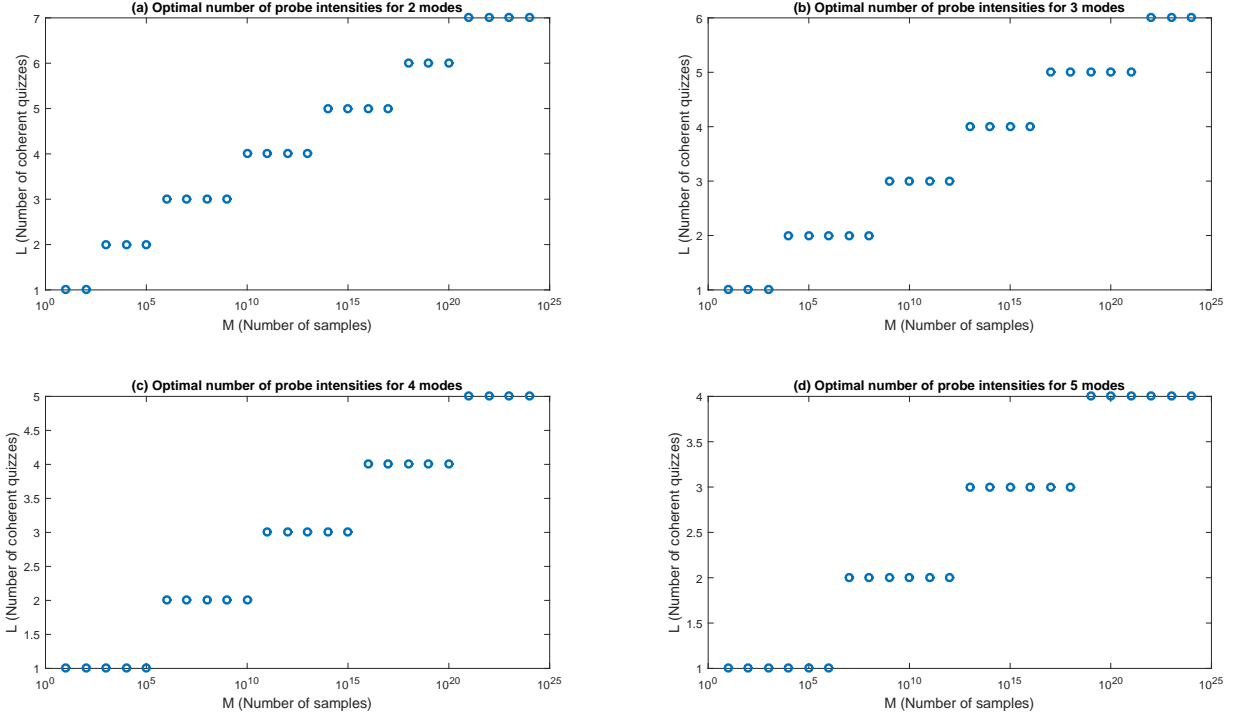}}
\caption{Optimized number of probe intensities for different input modes. }\label{Fig:optimalL}
\end{figure*}

Suppose the probe intensities for each mode are $\mu_j = j/L$ for $j = 0, 1, \dots, L$., then we have that
\begin{equation}\label{Eq:1}
\begin{aligned}
  \Delta_{t,n} &= \sqrt{\frac{(L+1)^n}{M}}f(L,1)^n + n\Delta_{L,1}. \\
  \end{aligned}
\end{equation}
Note that, we have $\log{\Delta_{L,1}} = -2.772 * L + 3.718$ and $\log{f(L,1)} = 0.67L + 0.189$, then
\begin{equation}\label{Eq:2}
\begin{aligned}
  \Delta_{t,n} &= \sqrt{\frac{(L+1)^n}{M}}e^{n(0.67L + 0.189)} + ne^{-2.772L + 3.718}. \\
  \end{aligned}
\end{equation}
We further optimize over $L$ to get a minimum total error of estimation, as shown in Fig.~\ref{Fig:TotalerrorParty}. The optimal number of probe intensities for different input modes are shown in Fig.~\ref{Fig:optimalL}

\section{Discussion}
In this work, we propose a way of simulating single photon with imperfect photon sources. We show that for a single photon, we can efficiently simulate it with coherent state. In addition, we generalize our result to multiple photon scenarios.

Our result indicate that small number of single photons can be well simulated by practical photon sources. In practice, this is useful for several information tasks. For instance, in quantum key distribution and quantum random number generation \cite{Cao16,YuanPhysRevA2015}, we can use phase randomized coherent states as source and at the same time guarantee the security.
In multipartite measurement device independent QKD \cite{Fu15}, our results can be applied to increase the key rate. In computation tasks, such as boson sampling, we can simulate the circuit by inputting imperfect photon source. In measurement device independent entanglement witness for multipartite quantum states \cite{branciard2013measurement,Yuan14,Zhao16}, our method can also make use of imperfect photon source instead of single photon to witness multipartite entanglement. In general, our method can also be regarded as a simulator for general quantum computing circuits.

In our derivation, we take the phase randomized coherent states as an example. It is not hard to see that other practical photon sources with a different mixing of Fock states can also be used in our method. Different photon sources will have similar estimations and errors and may provide some further advantages for certain tasks.

\section*{Acknowledgments}
We acknowledge D.~Berry and Z.~Cao for the insightful discussions. This work was supported by the 1000 Youth Fellowship program in China and the NSERC Discovery Grant.

\appendix

\section{Vacuum + two probe intensities}\label{Sec:3probe}
In this case, Alice inputs three phase randomized coherent states. From the previous calculation, we know that the interval is minimized when one of the intensities is 0. By assuming that and using two other non-zero  intensities, $\mu,\nu$, we have three linear relations
\begin{equation} \label{SMCQC:one3LinEq}
\begin{aligned}
A_0 &= Y_0 \\
A_\mu &= Y_0 + \mu Y_1 + \frac{\mu^2}{2}Y_2 + \frac{\mu^3}{3!}Y_3 + \dots \\
A_\nu &= Y_0 + \nu Y_1 + \frac{\nu^2}{2}Y_2 + \frac{\nu^3}{3!}Y_3 + \dots
\end{aligned}
\end{equation}
First, eliminate $Y_0$ and define  $B_\mu = (A_\mu - Y_0)/\mu$ and $B_\nu = (A_\nu - Y_0)/\nu$, we get
\begin{equation} \label{SMCQC:one3LinEq1}
\begin{aligned}
\mu B_\mu &= \mu Y_1 + \frac{\mu^2}{2}Y_2 + \frac{\mu^3}{3!}Y_3 + \dots \\
\nu B_\nu &= \nu Y_1 + \frac{\nu^2}{2}Y_2 + \frac{\nu^2}{3!}Y_3 + \dots
\end{aligned}
\end{equation}
Then, we can eliminate $Y_2$,
\begin{equation}\label{}
\begin{aligned}
{\nu^{-1}B_\nu - \mu^{-1}B_\mu} &= \frac{\mu-\nu}{\mu\nu}Y_1 + \left(\frac{\nu}{3!}Y_3 + \frac{\nu^2}{4!}Y_4 + \dots\right)-  \left(\frac{\mu}{3!}Y_3 + \frac{\mu^2}{4!}Y_4 + \dots\right),
\end{aligned}
\end{equation}
and we can estimate $Y_1$ by
\begin{equation}\label{}
Y_1^{\mathrm{est}} = \mu\nu\frac{\nu^{-1} B_\nu - \mu^{-1} B_\mu}{\mu-\nu},
\end{equation}
and
\begin{equation}\label{}
Y_1=Y_1^{\mathrm{est}}  + \frac{\mu\nu}{\mu-\nu}\left(\frac{\mu-\nu}{3!}Y_3 +\frac{\mu^2-\nu^2}{4!}Y_4 + \dots\right).
\end{equation}
The estimation interval is given by the difference between the maximal and minimal values of $\frac{\mu\nu}{\mu-\nu}\left(\frac{\mu-\nu}{3!}Y_3 +\frac{\mu^2-\nu^2}{4!}Y_4 + \dots\right)$, that is
\begin{equation} \label{SMCQC:oneDelta2}
\begin{aligned}
\Delta_2 &= \frac{\mu\nu}{\mu-\nu}\left(\frac{e^\mu - 1 - \mu}{\mu^2} -\frac{e^\nu - 1 - \nu}{\nu^2} \right).
\end{aligned}
\end{equation}
For $\mu$ and $\nu$ being small, we can approximate $\Delta_2$ by
\begin{equation} \label{SMCQC:oneDelta22}
\begin{aligned}
\Delta_2 &= \frac{\mu\nu}{3!} +O\left[\frac{\mu\nu(\mu+\nu)}{4!}\right].
\end{aligned}
\end{equation}

\section{Deriving $Y_1^{est}$ and $\Delta_L$ for vacuum plus $L$ probe intensities} \label{App:LQuizzes}
Suppose $\mu_0=0$ and $\mu_1<\mu_2<\dots<\mu_L$, similar to Eq.~\eqref{SMCQC:one3LinEq}, the set of linear equations can be expressed according to,
\begin{equation} \label{SMCQC:oneSLinEq3}
\begin{aligned}
\begin{pmatrix}
  A_0 \\
  A_{\mu_1} \\
  A_{\mu_2} \\
  \vdots \\
  A_{\mu_L} \\
\end{pmatrix}
=
\begin{pmatrix}
  1 & 0 & 0 & 0 & \dots \\
  1 & \mu_1 & \frac{\mu_1^2}{2!} & \frac{\mu_1^3}{3!} & \dots \\
  1 & \mu_2 & \frac{\mu_2^2}{2!} & \frac{\mu_2^3}{3!} & \dots \\
  & & \vdots \\
  1 & \mu_L & \frac{\mu_L^2}{2!} & \frac{\mu_L^3}{3!} & \dots \\
\end{pmatrix}
\begin{pmatrix}
  Y_0 \\
  Y_1 \\
  Y_2 \\
  \vdots \\
  \\
\end{pmatrix}
\end{aligned}
\end{equation}
We can eliminate the vacuum term, by defining $B_{\mu_l}=\frac{A_{\mu_l}-A_0}{\mu_l}$, for $1\le l\le L$. Then the linear equations becomes,
\begin{equation} \label{SMCQC:oneSLinEq}
\begin{aligned}
\begin{pmatrix}
  B_{\mu_1} \\
  B_{\mu_2} \\
  \vdots \\
  B_{\mu_L} \\
\end{pmatrix}
=
\begin{pmatrix}
  1 & \mu_1 & \mu_1^2 & \dots \\
  1 & \mu_2 & \mu_2^2 & \dots \\
  & \vdots \\
  1 & \mu_L & \mu_L^2 & \dots \\
\end{pmatrix}
\begin{pmatrix}
  Y_1 \\
  Y_2/2! \\
  \vdots \\
 \\
\end{pmatrix}
\end{aligned}
\end{equation}

Define $\mathbf{A}=(B_{\mu_1}, B_{\mu_2}, \dots, B_{\mu_L})^\mathrm{T}$, $\mathbf{Y}=(Y_1, Y_2/2!, \dots, Y_L/L!,\dots)^\mathrm{T}$, and
\begin{equation} \label{Vandermonde}
\begin{aligned}\mathbf{V}=
\begin{pmatrix}
  1 & \mu_1 & \mu_1^2 & \dots \\
  1 & \mu_2 & \mu_2^2 & \dots \\
  & \vdots \\
  1 & \mu_L & \mu_L^2 & \dots \\
\end{pmatrix},
\end{aligned}
\end{equation}
Then, we can rewrite the linear equations by
\begin{equation}\label{Eq:AVY}
\mathbf{A} = \mathbf{V} \mathbf{Y},
\end{equation}
Define $V'$ to be the first $L$ columns of $V$,
\begin{equation} \label{Vandermonde2}
\begin{aligned}\mathbf{V'}=
\begin{pmatrix}
  1 & \mu_1 & \mu_1^2 & \dots \mu_1^{L-1}\\
  1 & \mu_2 & \mu_2^2 & \dots \mu_2^{L-1}\\
  & \vdots \\
  1 & \mu_L & \mu_L^2 & \dots \mu_L^{L-1}\\
\end{pmatrix},
\end{aligned}
\end{equation}
Then we can see that $\mathbf{V}'$ is a Vandermonde matrix. Denote the inverse of $\mathbf{V}'$ by $\mathbf{M}$, who element $M_{i,j}$ is given by \cite{macon1958inverses},
\begin{equation} \label{Eq:InverseVandermonde}
\begin{aligned}
M_{i,j}&=\frac{(-1)^{i-1}\sum_{1\le k_1<k_2<\dots<k_{L-i}\le L;k_1,k_2,\dots,k_{L-i}\ne j}\mu_{k_1}\mu_{k_2}\dots\mu_{k_{L-i}}}{\prod_{1\le l\le L;l\ne j}(\mu_l-\mu_j)}, \;\;\; \text{for} \;\; 1\le i<L \\
M_{n,j}&=\frac{1}{\prod_{1\le l\le L;l\ne j}(\mu_j-\mu_l)} \\
\end{aligned}
\end{equation}
Then we can multiply $\mathbf{M}$ for both sides of Eq.~\eqref{Eq:AVY} and get
\begin{equation}\label{}
  \mathbf{MA} = \mathbf{MV}\mathbf{Y}.
\end{equation}
That is,
\begin{equation} \label{SMCQC:oneSLinEq2}
\begin{aligned}M
\begin{pmatrix}
  B_{\mu_1} \\
  B_{\mu_2} \\
  \vdots \\
  B_{\mu_L} \\
\end{pmatrix}
=M
\begin{pmatrix}
  1 & \mu_1 & \mu_1^2 & \dots \\
  1 & \mu_2 & \mu_2^2 & \dots \\
  & \vdots \\
  1 & \mu_L & \mu_L^2 & \dots \\
\end{pmatrix}
\begin{pmatrix}
  Y_1 \\
  Y_2/2! \\
  \vdots \\
   \\
\end{pmatrix}
=
\begin{pmatrix}
  1 & 0 & 0 & \dots 0 & \sum_{1\le j\le L}M_{1,j}\mu_j^L &\dots\\
  0 & 1 & 0 & \dots 0 & \sum_{1\le j\le L}M_{2,j}\mu_j^L &\dots\\
  & \vdots \\
  0 & 0 & 0 & \dots 1 & \sum_{1\le j\le L}M_{L,j}\mu_j^L &\dots\\
\end{pmatrix}
\begin{pmatrix}
  Y_1 \\
  Y_2/2! \\
  \vdots \\
 \\
\end{pmatrix}
\end{aligned}
\end{equation}
By considering the first row, we have
\begin{equation} \label{first row}
\begin{aligned}
\sum_{1\le j\le L}M_{1,j}B_{\mu_j}&= Y_1+\sum _{k> L}\frac{Y_k}{k!}\sum_{1\le j\le L}M_{1,j}\mu_j^k ,\\
\end{aligned}
\end{equation}
where $M_{1,j}$ is given by Eq.~\eqref{Eq:InverseVandermonde},
\begin{equation} \label{Eq:M1x}
\begin{aligned}
M_{1,j}&=\prod_{1\le l\le L;l\ne j}\frac{\mu_l}{\mu_l-\mu_j}. \\
\end{aligned}
\end{equation}
Therefore, the estimation of $Y_1$ is given by
\begin{equation}\label{Eq:appestimate}
\begin{aligned}
Y_1^{\mathrm{est}} &= \sum_{1\le j\le L} M_{1,j}B_{\mu_j} \\
&= \sum_{1\le j\le L} \frac{A_{\mu_j}-A_0}{\mu_j} \prod_{1\le l\le L;l\ne j}\frac{\mu_l}{\mu_l-\mu_j}. \\
&= \mu_1\mu_2\dots\mu_L\sum_{j= 1}^{L}\frac{\mu_j^{-2} (A_{\mu_j}-A_0)}{\prod_{1\le l\le L;l\ne j}(\mu_l-\mu_j)} \\
\end{aligned}
\end{equation}

Define the remaining term by $R = \sum _{k> L}\frac{Y_k}{k!}\sum_{1\le j\le L}M_{1,j}\mu_j^k$, the estimation interval is given by the interval of the maximal and minimal possible value of the $R$
\begin{equation}\label{}
  \Delta_L = \max_{Y_k,\forall k>L}R - \min_{Y_k,\forall k>L}R.
\end{equation}
Denote $\alpha_{j}^k = M_{1,j}\mu_j^k$, that is
\begin{equation}\label{}
  \alpha_{j}^k = \mu_j^k\prod_{1\le l\le L;l\ne j}\frac{\mu_l}{\mu_l-\mu_j},
\end{equation}
then we can easily verify that (1) $\alpha_{j}^k$ is positive when $j$ is odd; (2) $|\alpha_{j}^k| < |\alpha_{j'}^k|$ when $j<j'$. Therefore, the term $\sum_{1\le j\le k}M_{1,j}\mu_j^k$ can be expressed as
\begin{equation}\label{}
  \sum_{1\le j\le L}M_{1,j}\mu_j^k = \sum_{1\le j\le L}(-1)^{j-1}|\alpha_{j}^k|.
\end{equation}
When $L$ is even, the sum can be grouped into $(|\alpha_{1}^k|-|\alpha_{2}^k|) + (|\alpha_{3}^k|-|\alpha_{4}^k|) + \dots (|\alpha_{L-1}^k|-|\alpha_{L}^k|)$ and we can see that $\sum_{1\le j\le L}M_{1,j}\mu_j^k$ is negative.
When $L$ is odd, the sum can be grouped into $|\alpha_{1}^k| + (-|\alpha_{2}^k|+|\alpha_{3}^k|) + (-|\alpha_{4}^k|+|\alpha_{5}^k|) + \dots (-|\alpha_{L-1}^k|+|\alpha_{L}^k|)$ and we can see that $\sum_{1\le j\le L}M_{1,j}\mu_j^k$ is positive. Therefore, the signs of $\sum_{1\le j\le L}M_{1,j}\mu_j^k$ are the same for a fixed $L$, i.e., $(-1)^{L+1}$. Consequently, the maximum (minimum) value of $R$ can be obtained when all of $Y_k$¡¯s are equal to the same value (either $0$ or $1$). We denote those two values as $R_{Y_k=0,\forall k>L}$ and $R_{Y_k=1,\forall k>L}$, respectively.

Define $R' = (-1)^{L+1}R$, then the estimation interval is given by
\begin{equation}\label{}
  \Delta_L = R'_{Y_k=1,\forall k>L} - R'_{Y_k=0,\forall k>L}.
\end{equation}
Note that $R'_{Y_k=0,\forall k>L} = 0$. To calculate $R'_{Y_k=1,\forall k>L}$, we know that $R$ only contains the $Y_{k}, \forall k>L$ terms and therefore, the values of the $Y_{j}, \forall j=0,1,\dots,L$ can not affect the value of $R$. To simplify the calculation of $A_{\mu_j}, \forall j=0,1,\dots,L$, we can consider the case where $Y_0=Y1=\dots=1$ and hence $A_{\mu_j}=e^{\mu_j}$. In this case, according to Eq.~\eqref{first row}, we have that
\begin{equation}\label{Eq:appDelta}
\begin{aligned}
  \Delta_L = R'_{Y_k=1,\forall k>L} &= (-1)^{L+1}\sum _{k> L}\frac{1}{k!}\sum_{1\le j\le L}M_{1,j}\mu_j^k\\
  & = (-1)^{L+1}(Y_1^{\mathrm{est}}-Y_1)\\
  & = (-1)^{L+1}\left(\mu_1\mu_2\dots\mu_L\sum_{j=1}^{L}\frac{\mu_j^{-2} (e^{\mu_j}-1)}{\prod_{1\le n\le L;n\ne j}(\mu_n-\mu_j)}-1\right)
\end{aligned}
\end{equation}

The estimation $Y_1^{\mathrm{est}}$ in Eq.~\eqref{Eq:appestimate} can be represented as a linear combination of $A_{\mu_j}$ as,
\begin{equation}\label{Eq:y1estimation}
\begin{aligned}
Y_1^{\textrm{est}} &=\sum_{j=1}^{\lceil L/2\rceil}\frac{\mu_1\mu_2\dots\mu_L\mu_{2j+1}^{-2} (A_{\mu_{2j+1}}-A_0)}{\prod_{1\le n\le L;n\ne j}(\mu_n-\mu_{2j+1})} - \sum_{j=1}^{\lfloor L/2\rfloor}\frac{-\mu_1\mu_2\dots\mu_L\mu_{2j}^{-2} (A_{\mu_{2j}}-A_0)}{\prod_{1\le n\le L;n\ne j}(\mu_n-\mu_{2j})}\\
&=\sum_{j=1}^{\lceil L/2\rceil} \lambda_{2j-1}(A_{\mu_{2j-1}}-A_0) - \sum_{j=1}^{\lfloor L/2\rfloor}\lambda_{2j}(A_{\mu_{2j}}-A_0)\\
&=\sum_{j=1}^{\lceil L/2\rceil} \lambda_{2j-1}A_{\mu_{2j-1}} - \sum_{j=1}^{\lfloor L/2\rfloor}\lambda_{2j}A_{\mu_{2j}}+\lambda_0A_0,
\end{aligned}
\end{equation}
where the  coefficients $\lambda_j$ are positive and given by
\begin{equation}\label{Eq:deflambda}
\begin{aligned}
\lambda_{0} &= \sum_{j=1}^L(-1)^j\lambda_j, \\
\lambda_{j} &= \frac{(-1)^{j+1}}{\mu_j}\prod_{1\le n\le L;n\ne j} \frac{\mu_n}{(\mu_n-\mu_j)}, \quad \text{for} \quad 1\le j\le L.
\end{aligned}
\end{equation}
Similarly, the estimation error $\Delta_L$ is given by
\begin{equation}\label{SMCQC:deltaLestimation}
\begin{aligned}
\Delta_L  &=(-1)^{L+1}\left(\sum_{j=1}^{\lceil L/2\rceil} \lambda_{2j-1}e^{\mu_{2j-1}} - \sum_{j=1}^{\lfloor L/2\rfloor}\lambda_{2j}e^{\mu_{2j}}+\lambda_0-1\right),
\end{aligned}
\end{equation}

\bibliographystyle{apsrev4-1}

\bibliography{BibCQC}

\end{document}